\documentclass[pop,fleqn]{w-art}
\usepackage{times}
\usepackage{w-thm}
\theoremstyle{plain}

\theoremstyle{definition}

\usepackage{graphicx}

\def\to{\rightarrow}
\def\be{\begin{equation}}
\def\ee{\end{equation}}
\def\bea{\begin{eqnarray}}
\def\eea{\end{eqnarray}}
\def\nonu{\nonumber \\{}}

\def\a{\alpha}

\def\e{\epsilon}
\def\f{\phi}

\def\k{\kappa}

\def\o{\omega}
\def\p{\pi}

\def\s{\sigma}

\def\x{\xi}

\def\G{\Gamma}

\def\S{\Sigma}

\begin{document}
%%    The information for the title page will be placed between
%%    \begin{document} and \maketitle. The order of most entries
%%    is determined by the class file and can not be changed by
%%    rearranging them. The maketitle command follows after the
%%    abstract.
%%
%%    Most of the following commands will be completed by the publisher.
%%
%%    The copyrightyear is defined in the .clo file as the first argument
%%    of the copyrightinfo command. If the copyrightyear differs from that
%%    value it might be adjusted by the following definition:
%%
%% \renewcommand{\copyrightyear}{2003}% uncomment to change the copyrightyear.
%%
%\DOIsuffix{theDOIsuffix}
%%
%% issueinfo for header and copyright line
%\Volume{51}
%\Issue{1}
%\Month{01}
%\Year{2003}
%%
%%    First and last pagenumber of the article. If the option
%%    'autolastpage' is set (default) the second argument may be left empty.
%\pagespan{3}{}
%%
%%    Dates will be filled in by the publisher. The 'reviseddate' and
%%    'dateposted' (Published online) entry may be left empty.
%\Receiveddate{}
%\Reviseddate{30 November 2003}
%\Accepteddate{}
%\Dateposted{3 December 2004}
%%
%\keywords{Black Holes in String Theory, D-branes} %\subjclass[pacs]{04A25}

%% \pretitle{Editor's Choice}

%% We have a short and a long form for the title. The short form
%% (optional argument) goes into the running head.

\title[Microstates and near-horizon D-brane probes]{Microstates and near-horizon D-brane probes}

%% Please do not enter footnotes or \inst{}-notes into the optional
%% argument of the author command. The optional argument will go into
%% the header.  If there is only one address the marker \inst{x} may be
%% omitted.

%% Information for the first author.
\author[J. Raeymaekers]{Joris Raeymaekers%\inst{University of Tokyo}
  \footnote{E-mail:~\textsf{joris@hep-th.phys.s.u-tokyo.ac.jp}
  %,
  %          Phone: +00\,999\,999\,999,
  %          Fax: +00\,999\,999\,999
  }}
\address[]{Department of Physics, University of Tokyo,
Hongo 7-3-1, Bunkyo-ku, Tokyo 113-0033, Japan}
%%
%%    Information for the second author
%\author[S. Author]{Second Author\inst{1,2,}\footnote{Second author footnote.}}
%\address[\inst{2}]{Second address}
%%
%%    Information for the third author
%\author[T. Author]{Third Author\inst{2,}\footnote{Third author footnote.}}
%%
%%    \dedicatory{This is a dedicatory.}
\begin{abstract}
In \cite{Gaiotto:2004ij}, Gaiotto, Strominger and Yin  proposed a novel way
of counting black hole microstates by counting the quantum
mechanical ground states of probe branes placed in the
near-horizon black hole background. We  discuss the  generalization of this proposal
 to the case of two-charge
D0-D4 `small' black holes in type IIA. We also describe the
construction of BPS D-brane probes in the near-horizon region of
the 2-charge D1-D5 system in type IIB. Based on
\cite{Kim:2005yb},\cite{Raeymaekers:2006np}.
\end{abstract}
%% maketitle must follow the abstract.
\maketitle                   % Produces the title.

%% If there is not enough space inside the running head
%% for all authors including the title you may provide
%% the leftmark in one of the following three forms:

%% \renewcommand{\leftmark}
%% {First Author: A Short Title}

%% \renewcommand{\leftmark}
%% {First Author and Second Author: A Short Title}

%% \renewcommand{\leftmark}
%% {First Author et al.: A Short Title}

%% \tableofcontents  % Produces the table of contents.

\section{Introduction}
String theory has provided a microscopic understanding of large a class of supersymmetric black holes.
%Microscopic entropy calculations typically involve taking a suitable decoupling limit, relying heavily on supersymmetry
%protection under taking this limit.
In \cite{Gaiotto:2004ij}, Gaiotto, Strominger and Yin (GSY) proposed a novel way
of counting black hole microstates by counting the quantum
mechanical ground states of probe branes placed in the
near-horizon black hole background.

GSY considered a class of black holes in type IIA on $CY_3$
carrying D0 charge $q_0$ and charges $p^A$ from
D4-branes wrapping a generic 4-cycle in the Calabi-Yau space.
The near-horizon region is an   $AdS_2 \times S^2 \times CY_3$ attractor geometry \cite{Ferrara:1995ih}.
They considered the quantum mechanics living on D0-branes placed in this attractor background.
The super-isometry group of the background
 acts as a superconformal symmetry group on the quantum mechanics \cite{Gaiotto:2004pc}.
 It was proposed that the black hole microstates should be identified with
 the chiral primaries of this superconformal quantum mechanics. This can be seen as a concrete proposal for an
 $AdS_2/CFT_1$ duality.

An important ingredient in implementing this proposal is the property that $N$ D0-branes in the attractor background
can clump together to form nonabelian bound state configurations through a form of the Myers effect.
For sufficiently large $N$, these can be described as D2-branes
wrapping the $S^2$ and carrying $N$ units of worldvolume magnetic flux \cite{Simons:2004nm}.
They are static with respect to the global $AdS_2$ time coordinate and the corresponding Hamiltonian has a
discrete bound-state spectrum.

Such probe branes
experience a magnetic field along the Calabi-Yau directions induced  by the Wess-Zumino coupling $\int C^{(3)}$ to
the  D4-branes in the background.
The chiral primary states were shown to be in one-to-one correspondence to lowest Landau levels in this magnetic field,
and their degeneracy was found to
exactly reproduce the leading order entropy formula. However, this counting does not capture
the corrections to the entropy formula  subleading in
the D4-charges $p^A$.

When some of the D4-charges $p^A$ are taken to zero, 2-cycles in the attractor geometry shrink to zero size and the
above analysis breaks down due to the fact that higher derivative corrections to the supergravity action become important.
It is therefore a nontrivial question whether a GSY-inspired approach can still account for the black hole entropy
of such black holes.
 We will consider here
two examples of 2-charge `small' black holes, carrying D0-charge and only one type of D4-brane charge.
The examples discussed  here preserve a large amount of supersymmetry, which is presumably the
reason for their tractability.
We will show that the GSY proposal correctly accounts for the black hole entropy in these examples \cite{Kim:2005yb}.

It would also be of great interest to generalize the GSY proposal to account for the microstates of other black holes
or D-brane systems. The first step in such a program is the identification of suitable BPS probe branes in the near-horizon
geometry. In the second part of this note, we will consider the 2-charge D1-D5 system in type IIB, forming a black string in
6 dimensions with near-horizon $AdS_3 \times S^3$ geometry. We give a classification of half-BPS probe branes that
are string-like in $AdS_3$ and discuss some of their properties \cite{Raeymaekers:2006np}.

The results discussed here have appeared in \cite{Kim:2005yb,Raeymaekers:2006np}, to which we
refer for details and a more complete list of references.

\section{Superconformal quantum mechanics of small black holes}

The analysis of GSY in \cite{Gaiotto:2004ij} was performed for `large' black holes, which
have a nonvanishing horizon area in the leading supergravity
approximation. This is the case if the D4-brane charges $p^A$ are chosen such that
$ C_{ABC} p^A p^B p^C \neq 0 $ where $C_{ABC}$ are the triple intersection numbers on $CY_3$.
Furthermore, all $p^A$ have to be taken to be nonvanishing and
large in order for $\a'$ corrections to the background to be
suppressed. When $C_{ABC} p^A p^B p^C =  0$, the horizon area vanishes in the supergravity approximation and
higher derivative corrections cannot be neglected. On general grounds,
a horizon is expected to appear once these corrections are taken into account \cite{Sen:1995in},
hence these objects are called `small' black holes. We will address the GSY-inspired microstate counting in two examples.

As a first example, we consider type IIA compactified on $ T^2 \times K_3$. The 4-dimensional effective theory is an $N=4$
 supergravity. The 2-charge system of interest consists of $q_0$ D0-branes and
 $p^1$ D4-branes, the latter wrapped on $K_3$. Such configurations are half-BPS
and have a heterotic dual microscopic description as BPS excitations in the fundamental
string spectrum, the Dabholkar-Harvey states \cite{Dabholkar:1989jt}.

In the supergravity approximation,
the corresponding solution has vanishing horizon area, but a
horizon is generated when one includes  the leading 1-loop
correction to the prepotential \cite{Dabholkar:2004yr}.
The near-horizon geometry is
determined in terms of the charges by   generalized attractor
equations \cite{LopesCardoso:2000qm}. The resulting ten-dimensional IIA background is
$AdS_2 \times S^2 \times T^2 \times K_3$ with nonzero 2-form and 4-form RR flux:
\bea ds^2 &=& R^2 \left( - r^2 {dt^2} +
{dr^2 \over r^2}  + d \theta^2 + \sin^2 \theta d \f^2 \right)
+ 2 dz d \bar z + 2 r g_{a \bar b} d z^a d \bar z^{\bar b}\nonu
F^{(4)} &=&  {p^1 \over 4 \p} \sin \theta d \theta d \f \wedge
\o_1; \qquad F^{(2)} =  {R \over g_s} dr \wedge dt
 \label{nearhor10d}
\eea Here, we have chosen coordinates $(z, \bar z)$ on $T^2$ and
$(z^a, \bar z^{\bar a})_{a, \bar a=1,2} $ on $K_3$, $g_{a \bar
b}$ is proportional to the asymptotic Ricci-flat metric on $K_3$, and
$\o_1$ is the normalized volume form on $T^2$.
In units in which $2 \p \sqrt{\a '} =1$, the radius $R$ of $AdS_2
\times S^2$ is given by $R = {g_s \over 2 \p} \sqrt{ p^1 \over
q_0}$. It's important to note that the volume  of $K_3$ is
not fixed to a  finite value at the horizon but varies  like
$r^2$. This is a consequence of the fact there are no D4-branes
wrapped on the cycles dual to the 2-cycles in $K_3$, hence the
size of these cycles is not fixed by the attractor mechanism. This
constitutes an important difference with the large black holes
studied in \cite{Gaiotto:2004ij}, where all internal four-cycles
have a large number of D4-branes wrapped on them.

As in  \cite{Gaiotto:2004ij}, we will consider the quantum
mechanics of a nonabelian configuration of $N$  D0-brane probes in
the background (\ref{nearhor10d}) forming a fuzzy sphere of radius $R$.
 For $R^2 / N \ll 1 $, this system has an
equivalent description in terms of a D2-brane wrapping the $S^2$
with $N$ units of flux turned on on its worldvolume. The terms contributing to
the bosonic worldvolume action are
$$ S = T_2 \int d^3 \s e^{-\phi} \sqrt{ - \det ( G + F ) }  + T_2 \int_{D2} C^{(3)} + T_2
\int_{D2} F\wedge C^{(1)}$$
The bosonic Hamiltonian, to quadratic order in
derivatives and in the limit $R^2 / N \ll 1$  is
\be  H = {1\over 8RT} P_\x^2 + {R \over T \x^2} (P_z - A_z)
(P_{\bar z} - A_{\bar z}) + {32 \p^4 R^5 \over g_s^2 N \x^2}
 + {R \over T} P_a g^{a \bar b} P_{\bar b}\label{bosham}\ee
where $\x = 1/\sqrt{r},\ T = {2 \p \over g_s} \sqrt{(4 \p R^2)^2 + N^2}$ and we have introduced
a $U(1)$ gauge potential $A$ on $T^2$
obeying $ d A = 2 \p p^1 \o_1$. We denoted the canonical momenta conjugate to $z, z^a$ by $P_z, P_a$ respectively.
One notes that the Hamiltonian (\ref{bosham}) is a sum of two decoupled parts: the first three terms
describing the dynamics on ${\bf R} \times T^2$ (the ${\bf R}$ representing the  the radial $AdS_2$ direction),
and the last term describing the motion on
$K_3$.
This decoupling is a direct consequence of the radial dependence of the $K_3$ volume modulus in (\ref{nearhor10d}).
The full  quantum mechanics also contains sixteen fermions, which we have not displayed here.

The (super-)isometries of the background (\ref{nearhor10d}) act as
symmetries on the quantum mechanics, giving the symmetry algebra a
superconformal structure. Due to the decoupling of the ${\bf
R}\times T^2$ and $K_3$ parts of the Hamiltonian, the symmetry
group naturally splits into  a group acting on the ${\bf R}\times
T^2$ and $K_3$ parts of the wavefunction respectively. It turns
out that the symmetry algebra of the  ${\bf R}\times T^2$ part is
the $N=4$ superconformal algebra $su(1,1|2)_Z$, where $Z$
indicates the presence of a central charge. It contains the
conformal algebra $sl(2,R)$, an $su(2)$ R-symmetry and and 8
fermionic generators. The motion on $K_3$ has the structure of an
$N=4$ supersymmetric quantum mechanics (SQM).
 The
GSY proposal made in \cite{Gaiotto:2004ij} states that the chiral
primaries of the near-horizon D0-brane quantum mechanics are to be
identified with the black hole microstates. In the case at
hand, we should tensor the chiral primaries of $su(1,1|2)_Z$ with
with the supersymmetric ground states of the $N=4$ SQM.

As in \cite{Gaiotto:2004ij}, the chiral primaries of $su(1,1|2)_Z$  can be shown to be in
one-to-one correspondence with lowest Landau level wavefunctions for a particle
moving on $T^2$ in the presence of the magnetic field $A$. The number of independent
lowest Landau level wavefunctions is given by an index theorem and
is equal to the first Chern number
${1 \over 2 \p}  \int_{T^2} d A = p^1.$
These states should be tensored with supersymmetric ground states of the $N=4$ SQM on $K_3$.
A standard construction maps the  supersymmetric ground states of $N=4$ SQM on any K\"{a}hler manifold
to Dolbeault cohomology classes, the even and odd forms corresponding to bosons and fermions
respectively. Hence on $K_3$ we have an $N=4$ SQM with 24 bosonic supersymmetric ground states.

Tensoring those together we find a total of $24 p^1$
bosonic chiral primaries. Since the number of ground states doesn't depend on the background D0-charge $q_0$,
one can take
$q_0 \to 0$ so that all of the D0 charge comes  from the probes and is equal to  $N$.
There is a large
degeneracy of  states coming from the many ways the  total number of
D0-branes charge $N$  can be partitioned into smaller clusters,
each cluster corresponding to a wrapped D2-brane that can reside
in any of the $24 p^1$ chiral primaries. The ground state degeneracies $d_N$
can be summarized in a
generating function
$$ Z = \sum_N d_N q^N = \prod_n ( 1-
q^n)^{-24 p^1}.$$
This gives the asymptotic
degeneracy at large $N$
$$
\ln d_N \approx 4 \p \sqrt{N p^1}
$$
which indeed equals  the known asymptotic degeneracy obtained
from microscopic counting \cite{Dabholkar:1989jt} or from the
supergravity description incorporating higher derivative
corrections \cite{Dabholkar:2004yr}. We note that the known subleading corrections
to the entropy are not captured by the above partition function, and their incorporation in this
framework
remains an open problem.

The above analysis can be repeated for $K_3$ replaced by a four-torus $T^4$, under some additional assumptions.
The corresponding small black hole is
$1/4$ BPS in the effective  4-dimensional $N=8$ supergravity.
Since
all corrections to the prepotential vanish in this case, the
corrections that generate the horizon are expected to come from
non-holomorphic corrections to the supergravity equations, and it
is not known how to incorporate these systematically at present.
We shall be  cavalier  and simply  assume that the
near-horizon limit of the corrected background is still of
the form (\ref{nearhor10d}), with the $K_3$ metric now replaced by
the flat metric on $T^4$ and possibly with a different value of the
constant $R$.
The  above analysis can then be repeated,
the only difference coming from the counting of ground states  of the
$N=4$ SQM, now corresponding to the Dolbeault cohomology of $T^4$.
This gives  8
bosonic and 8 fermionic ground states. The
partition function is now
$$ Z = \prod_n \left( {1 + q^n \over 1 -
q^n} \right)^{8 p^1}. \label{partT6}$$
This gives the asymptotic
degeneracy
$$ \ln d_N \approx 2 \sqrt{2} \p \sqrt{N p^1} $$
which is in agreement with the known degeneracy from microscopic
counting \cite{Dabholkar:1989jt}.

\section{Supersymmetric D-branes in the D1-D5 background}

In order to put the GSY proposal on a sounder footing and address some of its difficulties, it would be of
great interest to extend it to other black holes
or D-brane systems. Here, we will consider the well-studied 2-charge D1-D5 system in type
IIB compactified on $M$ (where $M$ can be $T^4$ or $K_3$), forming
a black string in 6 dimensions. The first step  is the identification of suitable BPS probe branes in the near-horizon
geometry which is $AdS_3\times S^3 \times M$. Even though the D1-D5 system is T-dual to the small D0-D4 black holes
considered above,
T-duality does not commute with taking the near-horizon limit, making a direct mapping between near-horizon
microstates difficult.
We will instead look directly for supersymmetric D-branes in the near horizon  region
with  properties similar to the near-horizon microstates of the D0-D4 small black holes.

The near-horizon geometry reads, in Poincar\'{e} coordinates:
\bea
ds^2 &=& r_1 r_5 [ u^2 (-dt^2+dx^2) + {du^2 \over u^2}
  + d \psi^2 + \sin^2 \psi (d \theta^2 + \sin^2 \theta d \phi^2) ] +
{r_1 \over r_5} ds^2_{M}\nonu
%e^{-\f} &=& {1 \over g} {r_5 \over  r_1}\nonu
F^{(3)} &=& {2 r_5^2 \over g} [ u  dt \wedge dx \wedge du
+ \sin^2 \psi \sin \theta  d\psi \wedge d \theta \wedge d \phi ]; \qquad e^{-\f} = {1 \over g} {r_5 \over  r_1} \nonumber
\eea
We restrict attention to branes that are string-like in $AdS_3$ and span an $AdS_2$ subspace. These can  be
 parameterized as
$$ u = {C \over x}$$
for some constant $C$. Such branes
are static with respect to the global time coordinate, and furthermore lead to good open string boundary
conditions \cite{Bachas:2000fr}. We allow the branes to extend in the compact directions and
 carry arbitrary worldvolume
fluxes (and hence also induced lower-dimensional D-brane charges).

The near-horizon geometry preserves 16 supersymmetries, which split into 8 `Poincar\'{e}' supersymmetries that extend
to the full asymptotically flat solution
and  8 `conformal' supersymmetries that exist only in the near-horizon limit.
The condition for a D-brane probe to preserve some supersymmetry can be written as \cite{Bergshoeff:1997kr}
$$ (1 - \G ) \e = 0$$
where
  $\G$ (satisfying ${\rm tr} \G = 0,\ \G^2 = 1$) is the operator entering in the $\k$-symmetry
transformation rule on the D-brane and depends on the brane embedding as well as the worldvolume gauge fields;
$\e$  are the Killing spinors of the background pulled back to the
world-volume.

Referring to \cite{Raeymaekers:2006np} for calculational details, one finds
in this manner a  large variety of D-branes preserving half of the
16 near-horizon supersymmetries. All of them turn out to preserve  half  of 8 the `Poincar\'{e}'
supersymmetries as well. They are summarized in the  following table which lists
the submanifold spanned by the brane as well as possible restrictions on the embedding and/or worldvolume gauge fields.
\begin{center}
\begin{tabular}{|c|c|c|c|c|}\hline
brane  & $AdS_3$ & $S^3$ & $M$ & restrictions    \\ \hline \hline
D1 & $AdS_2$ & $\cdot$ & $\cdot$ &   \\ \hline D3 & $AdS_2$ &
$\cdot$ & 2-cycle $\S$ & $\S$ holomorphic  \\ \hline D5 & $AdS_2$
& $\cdot$ & $M$ &   \\ \hline \hline D3 & $AdS_2$ & $S^2$ &
$\cdot$ &  \\ \hline D7 & $AdS_2$ & $S^2$ & $M$ &  $F_{|M}$
antiselfdual \\  \hline \hline
\end{tabular}
\end{center}
The solutions come in two types: branes of the first type are pointlike on the $S^3$ while branes of the second type
wrap an $S^2$ within $S^3$. The latter
 are dipolar as the  $S^2$ is contractible  and is
stabilized by worldvolume
flux \cite{Bachas:2000ik}.

In the second category, let's discuss
the D3-branes spanning an $AdS_2 \times S^2$  in a little more detail (see also \cite{Bachas:2000fr,Pawelczyk:2000hy}).
The electric and magnetic worldvolume
fields induce fundamental string charge $q$ and D-string charge $p$. They can be seen as $(p,q)$ strings puffed
up to form a dipolar D3-brane through a form of the Myers effect. The size of the $S^2$ is related to the fundamental string charge
$q$ and its maximum value leads to an `exclusion bound' $q \leq Q_5$. This bound is similar to the upper bound on the angular momentum
for lowest Landau levels in the D0-D4 system discussed above, which is set by the background D4-charge. However, it is not
clear whether the $AdS_2 \times S^2$ branes are related to microstates of the D1-D5 system. Since they intersect the boundary of
$AdS_3$, they are not states in the dual CFT, but rather conformal defects as in \cite{Bachas:2001vj}. It would be interesting
to have a better understanding of these and the other branes in the above table from the point of view of the dual CFT.

%\begin{acknowledgement}
%  An acknowledgement may be placed at the end of the article.
%\end{acknowledgement}

\end{document}